\documentclass[12pt]{book}

\usepackage[dvips]{graphicx,color}
\usepackage{makeidx,tsukuba}

\makeauthorindex
\makeindex

\begin{document}

\BookTitle{\itshape The 28th International Cosmic Ray Conference}
\CopyRight{\copyright 2003 by Universal Academy Press, Inc.}
\pagenumbering{arabic}

\chapter{
Evaluation of Models for Diffuse Continuum Gamma Rays in EGRET Range}

\author{%
%
%
Andrew W.~Strong,$^1$ Igor V.~Moskalenko,$^{2,3}$ and Olaf Reimer$^4$ \\
{\it (1) Max-Planck Institut f\"ur extraterrestrische Physik, Garching, Germany\\
(2) NASA/Goddard Space Flight Center, Code 661, Greenbelt, MD 20771, USA\\
(3) JCA/University of Maryland, Baltimore County, Baltimore, MD 21250, USA\\
(4) Institut f\"ur Theoretische Physik IV, Ruhr-Universit\"at Bochum, Germany
} \\
}

\section*{Abstract}
The GALPROP model for cosmic-ray (CR) propagation produces explicit
predictions for the angular distribution of Galactic diffuse $\gamma$-rays. We compare
our current models with EGRET spectra for various regions of the sky.
This allows a critical test of alternative hypotheses for the observed
GeV excess. We show that a population of hard-spectrum $\gamma$-ray sources cannot
be solely responsible for the excess since it also appears at high
latitudes; on the other hand a hard CR electron spectrum model cannot
explain the $\gamma$-ray excess in the inner Galaxy. Hence some
combination of these explanations is suggested.

\section{Introduction}
Diffuse continuum $\gamma$-rays from the interstellar medium are potentially
able to reveal much about the sources and propagation of CR,
but in practice the exploitation of this well-known connection is
problematic. We have previously [3] compared a range of
models, based on our cosmic-ray propagation code GALPROP, with data
from the Compton Gamma Ray Observatory. While it is rather easy to get
agreement within a factor $\sim$ 2 from a few MeV to 10 GeV with a
``conventional'' set of parameters, the data quality warrant considerably
better fits. Specifically, in order to reproduce the excess at GeV
energies observed by EGRET, we adopted a model
with a rather hard electron injection spectrum and a proton spectrum
deviating moderately from that measured locally. 
An essential feature of our approach was that the locally-measured CR
spectrum of electrons is not a good constraint because of the spatial
fluctuations due to energy losses and the stochastic nature of the
sources in space and time; the average interstellar electron spectrum
responsible for $\gamma$-rays via IC emission can therefore
be quite different from that measured locally. The hard electron spectrum interpretation is not
entirely convincing however, since the electron fluctuations required
are larger than expected. In addition, new accurate measurements of the local
proton and Helium spectrum allow less freedom for deviations in the
$\pi^0$-decay component which were exploited in [3].
Another suggestion which has been made [1] is that the $\gamma$-ray
spectrum contains a $\pi^0$-decay component from CR protons close to their
(SNR) sources, so that in $\gamma$-rays we see the injection spectrum which is much
flatter than the propagated spectrum which we measure locally. 

In a companion paper [4] we use our model to re-determine the 
extragalactic $\gamma$-ray background (EGRB).

\section{Data and method}
We use the EGRET counts and exposure 
all-sky maps; 
the sources of the 3EG catalogue have been removed 
by the procedure described in [3].
The predicted model skymaps are convolved with the EGRET point-spread function.
The spectra are compared in the sky regions summarized in Table 1.
Region A corresponds to the
inner Galactic radian, region B is the Galactic plane excluding the inner radian, region C is the ``outer Galaxy'', regions D and E cover higher latitudes at all
longitudes, region F covers the Galactic poles. 
The models all
use the locally-observed proton and Helium spectra.
This is done because the nucleon data are now more precise 
(see references in [2]) than those which were available in [3].
The nucleon injection spectra and the
propagation parameters are chosen to reproduce the most recent
measurements of primary and secondary nuclei, as described in detail in
[2].
The halo height is taken as 4 kpc as in [3], in accordance with our analysis of CR
secondary/primary ratios [2].

%
\begin{table}[tb]
\caption{Sky regions used for comparison of models with data}
\begin{center}
\begin{tabular}{cccl}
\hline
Region &$l$, degrees &$|b|$, degrees &Description \\
\hline
A&330--30& 0--5 &Inner Galaxy \\
B&30--330& 0--5 &Galactic plane avoiding inner Galaxy \\
C&90--270& 0--10 &Outer Galaxy \\
D&0--360 & 10--20 &Intermediate latitudes 1 \\
E&0--360 & 20--60 &Intermediate latitudes 2 \\
F&0--360 & 60--90 &Galactic poles \\
\hline
\end{tabular}
\end{center}
\end{table}

\section{Conventional model}
We start by repeating the test of the ``conventional'' model which is
based on the locally-observed electron (as well as nucleon) spectrum.
The model spectra are compared with EGRET data in Fig.\ 1.
 As found in previous work, for such a model the GeV region shows an
excess relative to that predicted; what is now evident is that this
excess appears {\it in all latitudes/longitude ranges}.
This already
shows that {\it the GeV excess is not a feature restricted to the Galactic
ridge or the gas-related emission.}
Further it is clear that a simple upward rescaling of the $\pi^0$-decay
component will not improve the fit in any region, since the observed
peak is at higher energies than the $\pi^0$-decay peak. 
In the ``SNR source'' scenario [1] the spectrum in the inner Galaxy is
attributed to an additional population of unresolved SNR, but this
component cannot explain the excess at high latitudes, and hardly in
the outer Galaxy. 
This explanation is therefore by itself insufficient,
although as we will show it may well be part of the solution.

\section{Optimized model}
We next consider a model with a hard electron injection spectrum, as
presented in [3]. Our approach is to concentrate on
obtaining a fit in all regions {\it apart from the inner Galaxy}, since
then we can be reasonably sure that no population of unresolved sources is
distorting the spectrum, and individual sources would be nearby, rare and easily identified, and so the situation should be relatively simple. (A population of weak sources in the halo could also contribute, but this is beyond the scope of this investigation).
The result of our optimization procedure is shown in Fig.\ 2.
An electron injection spectral index of 1.9 (with no break) is found optimal, 
consistent with what was found in [3].
Values differing by up to 0.1 from this would also be acceptable.
In order to be consistent with EGRET data $>$10 GeV, a cutoff in the electron
spectrum at 3 TeV is required.
Overall the fit is satisfactory and much improved over the conventional
model 
and the GeV excess is reproduced in all regions except region A. 
At low latitudes in the inner Galaxy (region A) the peak around 1 GeV
 is not reproduced. Evidently a successful model requires
 {\it both} an adjustment of the IC via the electron spectrum {\it and}
 a hard-spectrum source population in the inner Galaxy.
The sources must be concentrated in the inner Galaxy since in the
remainder of the Galactic plane ($ 30^\circ<l<330^\circ, |b|<5^\circ$) the fit is satisfactory.
 Candidates for the source population include $\gamma$-ray pulsars like Geminga which does exhibit the required hard spectrum.

 \begin{figure}
  \begin{center}
\hskip2mm
\includegraphics[width=0.307\textwidth]{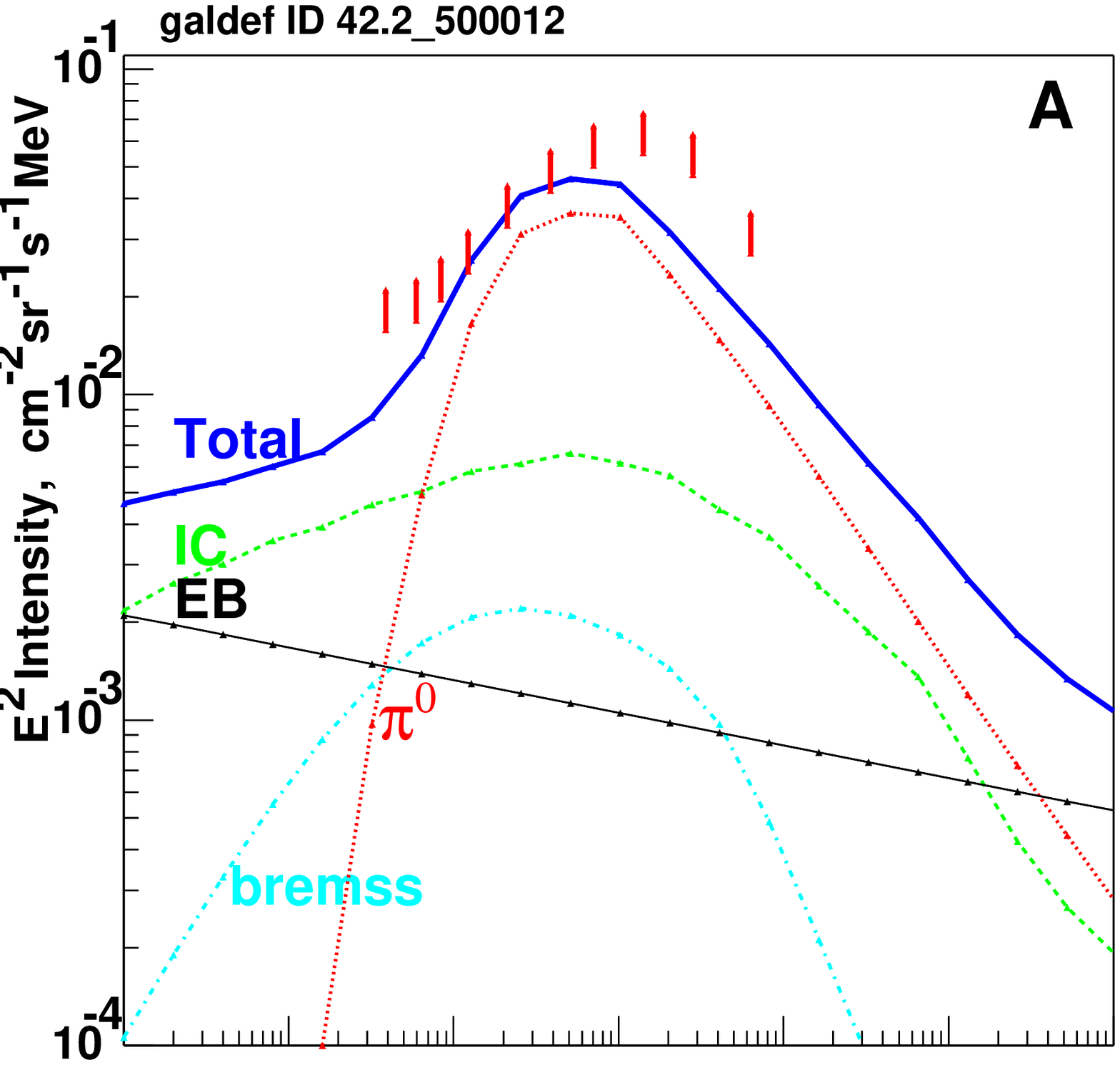}\hfill
\includegraphics[width=0.307\textwidth]{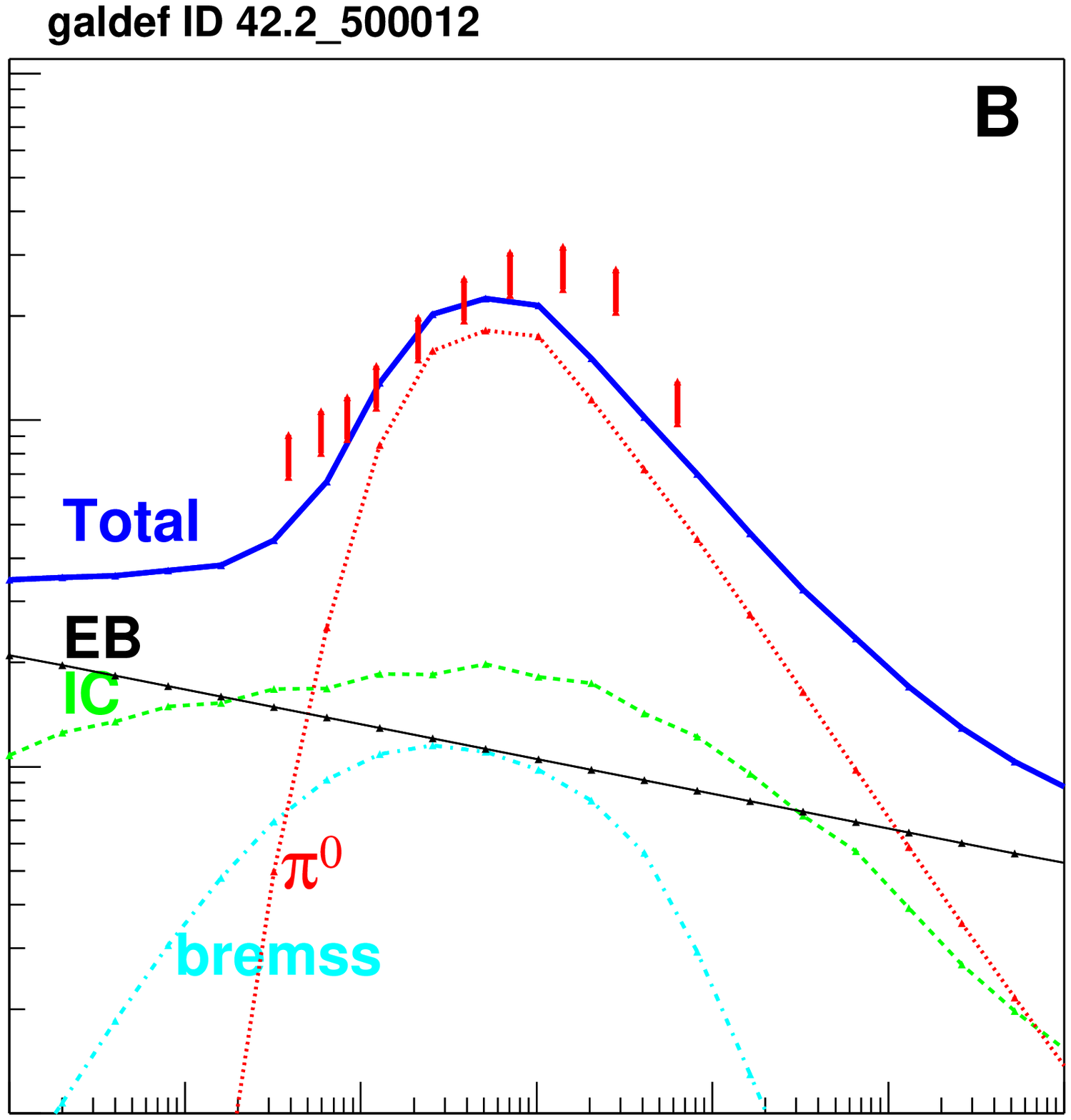}\hfill
\includegraphics[width=0.307\textwidth]{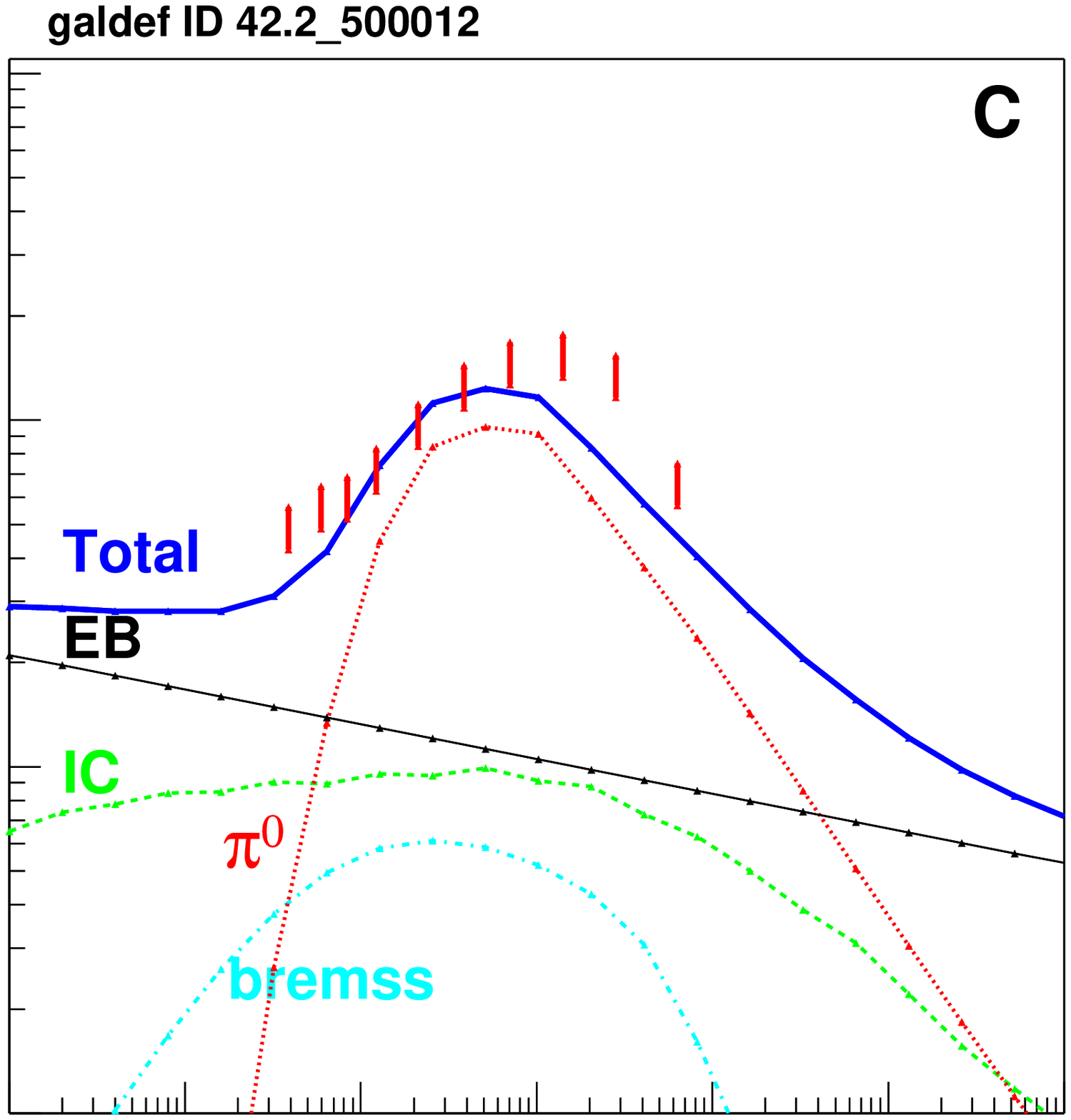}
\vskip5mm
\hskip2mm
\includegraphics[width=0.307\textwidth]{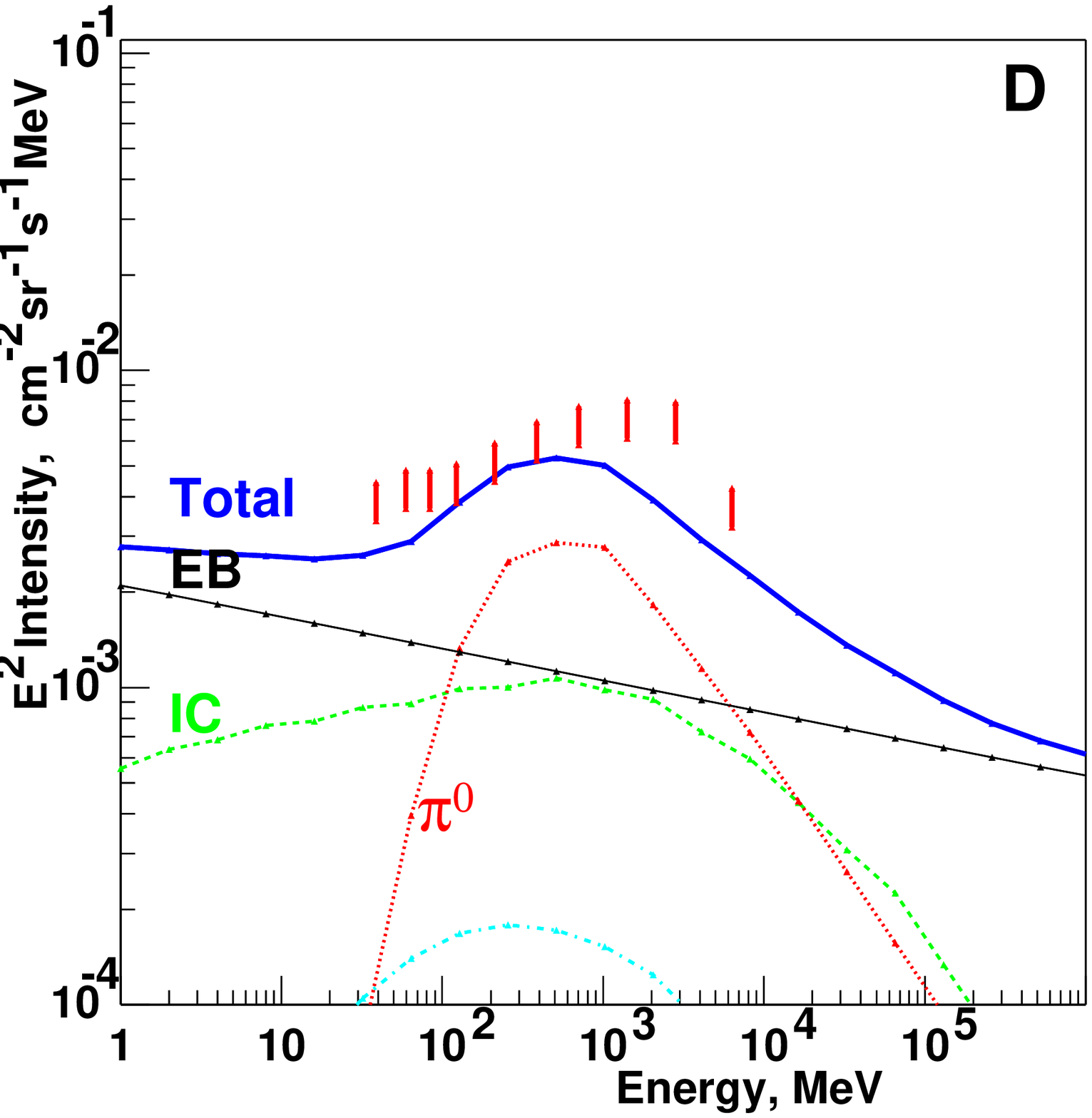}\hfill
\includegraphics[width=0.307\textwidth]{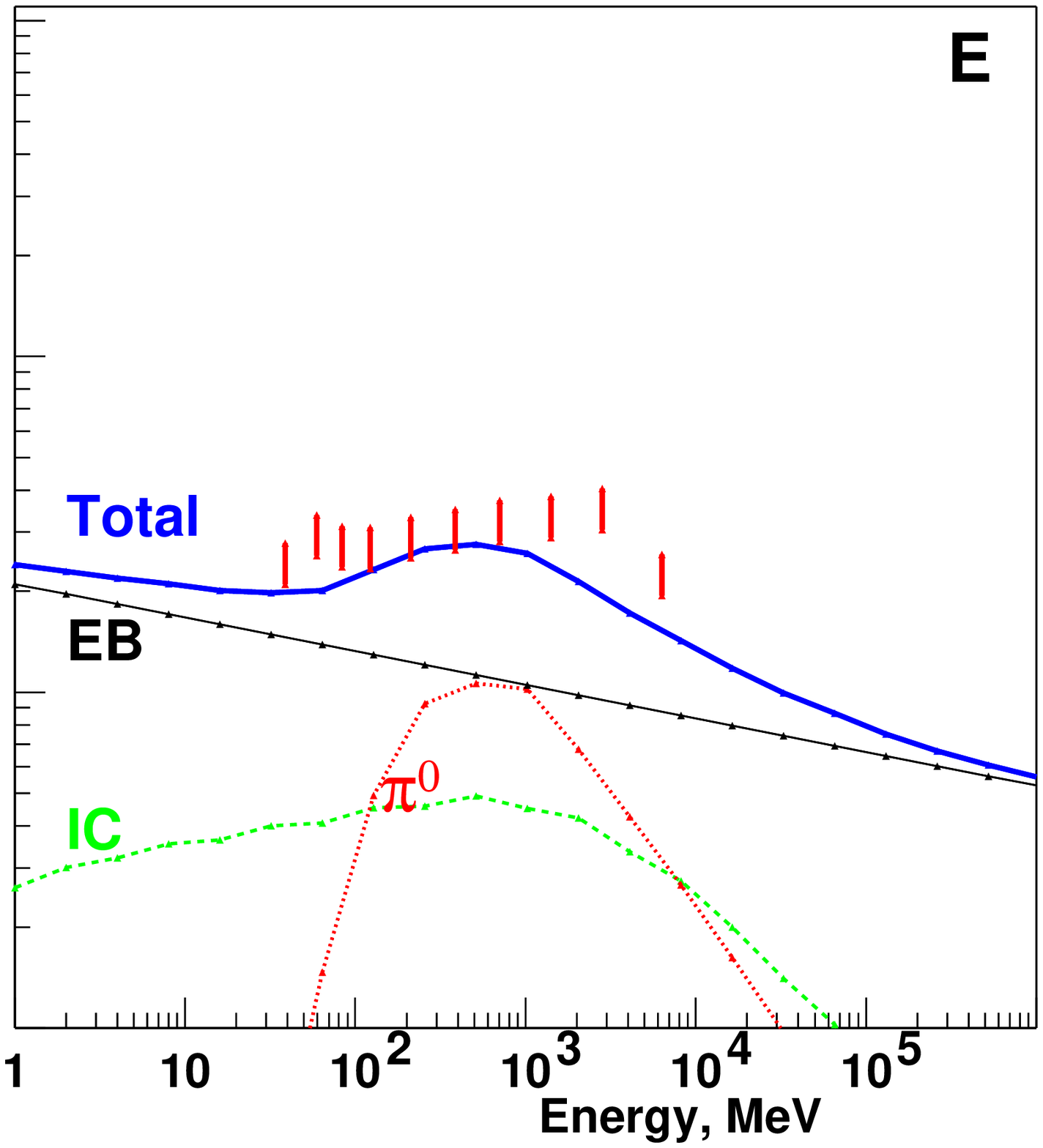}\hfill
\includegraphics[width=0.307\textwidth]{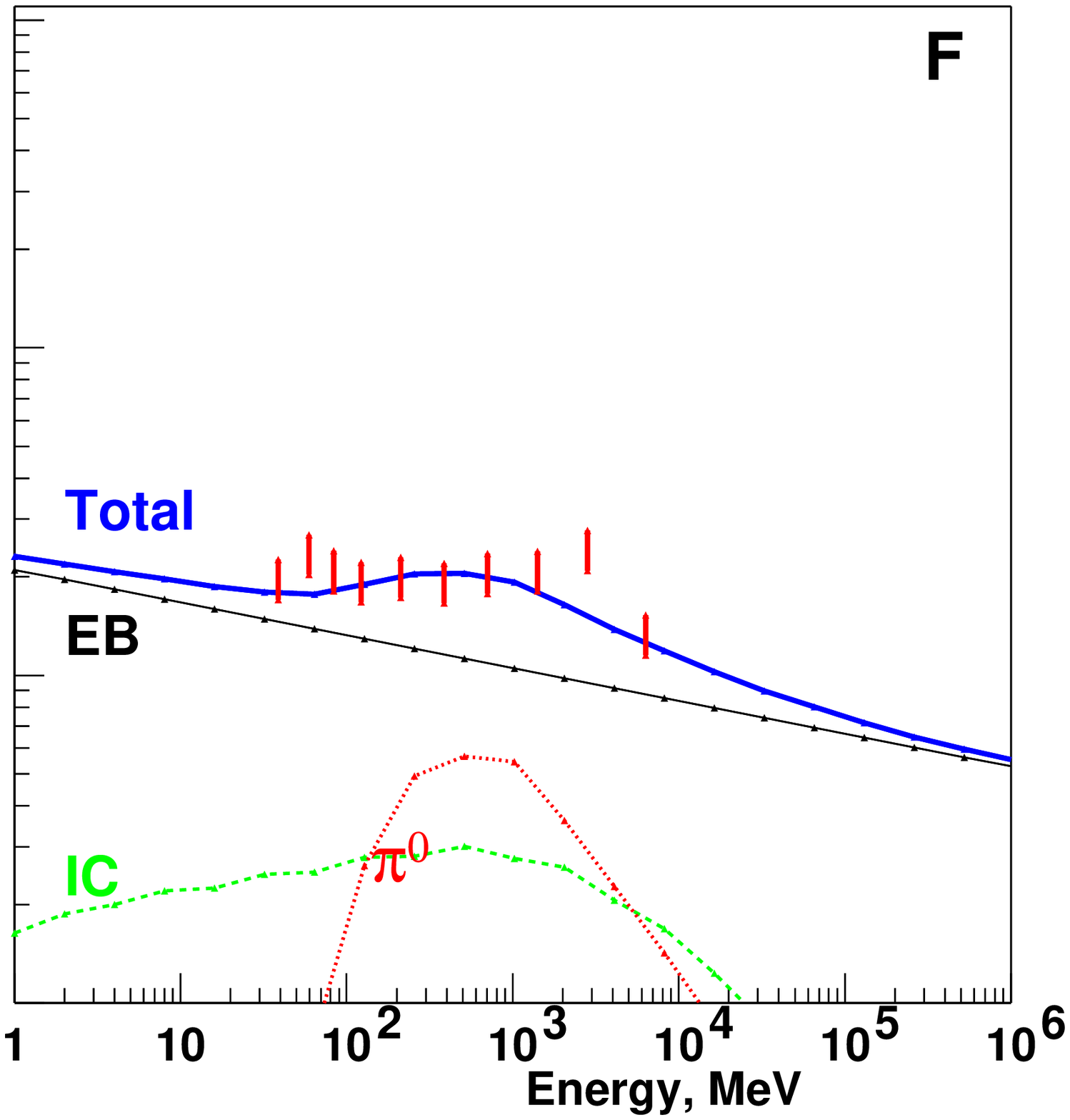}
  \end{center}
  \vspace{-0.5pc}

\caption{$\gamma$-ray spectra for conventional model for regions A -- F.} 
 \label{spectrum_500012}
 \end{figure}

 \begin{figure}
  \begin{center}
\hskip2mm
 \includegraphics[width=0.307\textwidth]{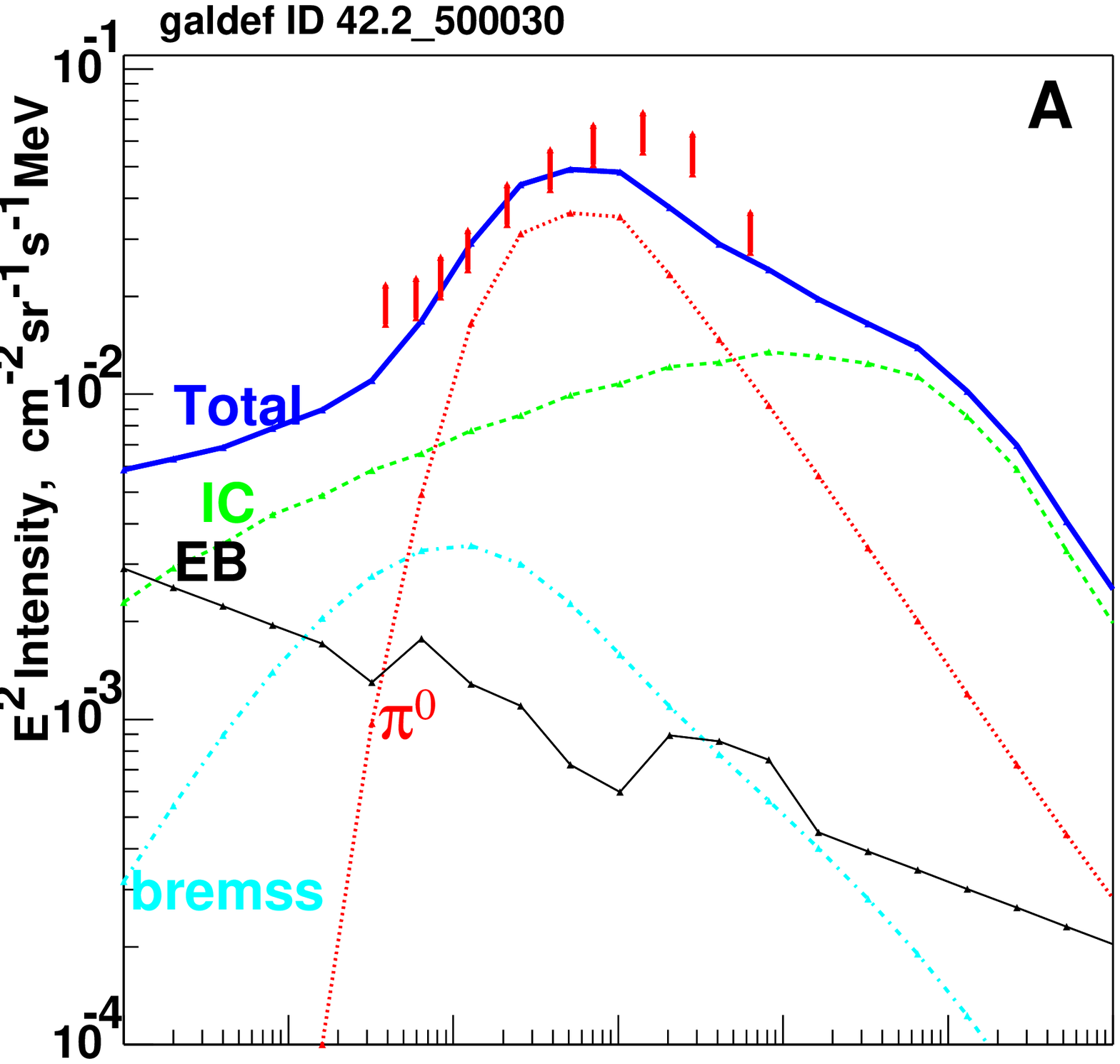}\hfill
 \includegraphics[width=0.307\textwidth]{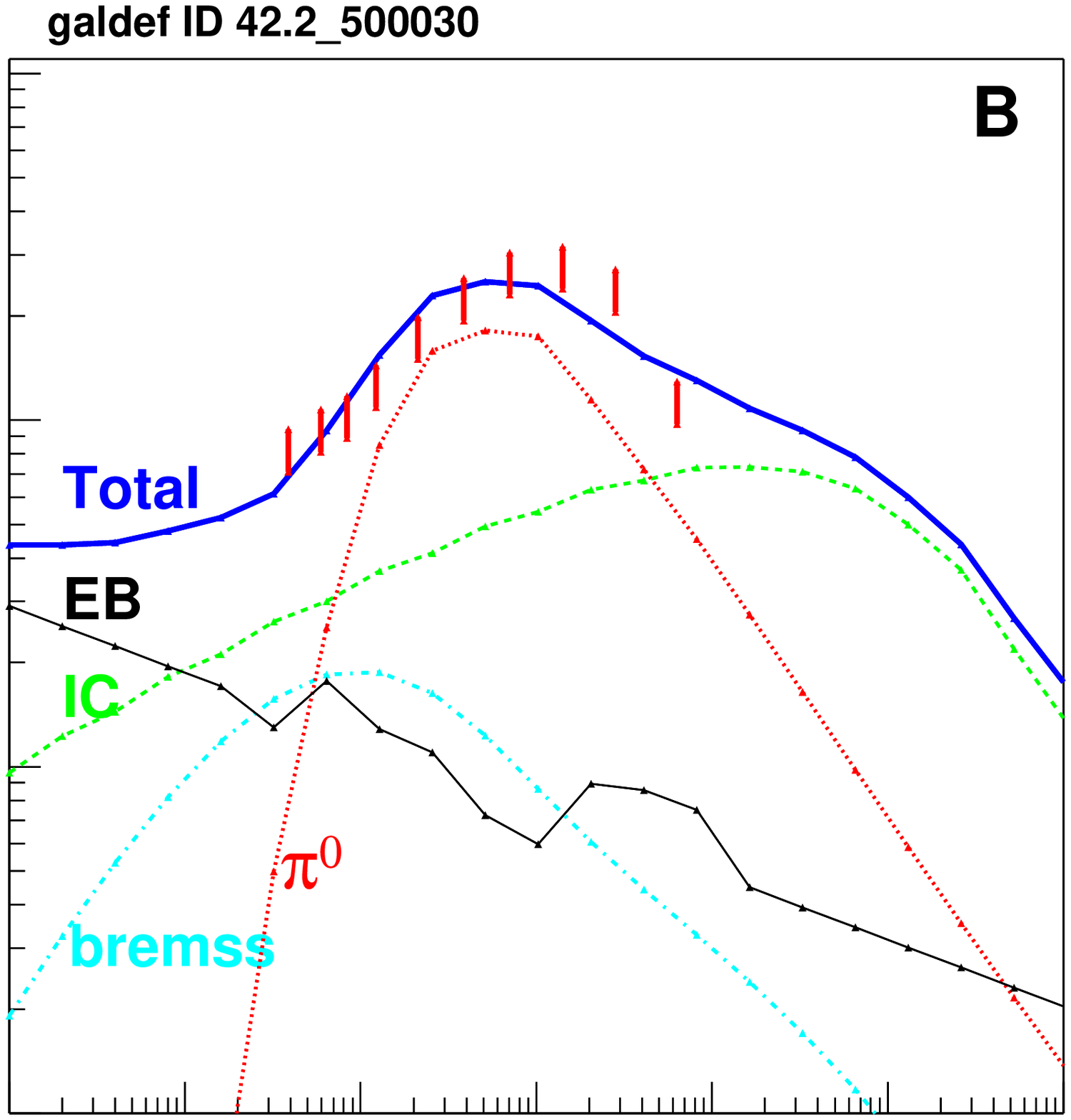}\hfill
 \includegraphics[width=0.307\textwidth,height=0.307\textwidth]{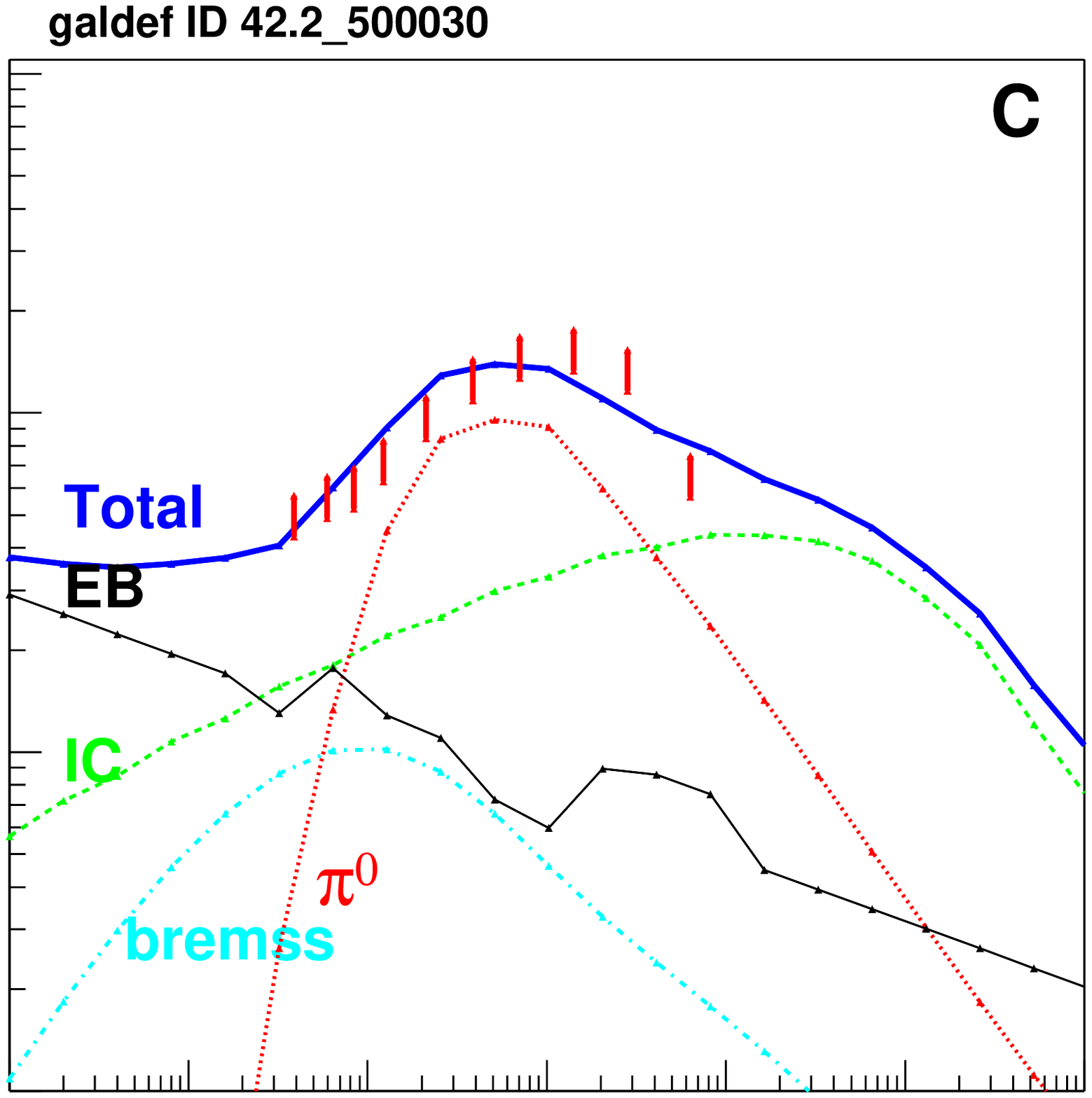}
\vskip5mm
\hskip2mm
 \includegraphics[width=0.307\textwidth,height=0.307\textwidth]{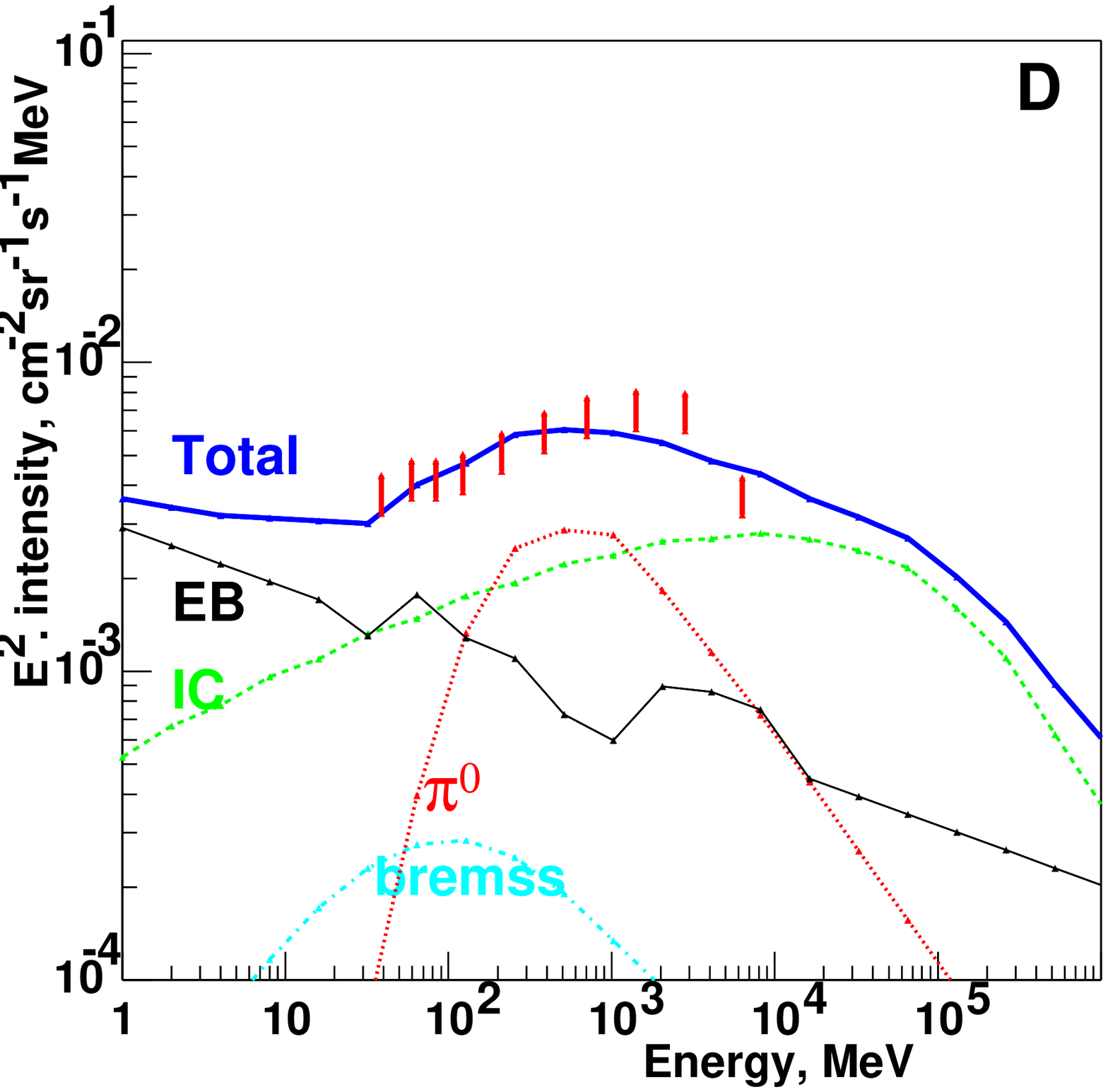}\hfill
 \includegraphics[width=0.307\textwidth,height=0.307\textwidth]{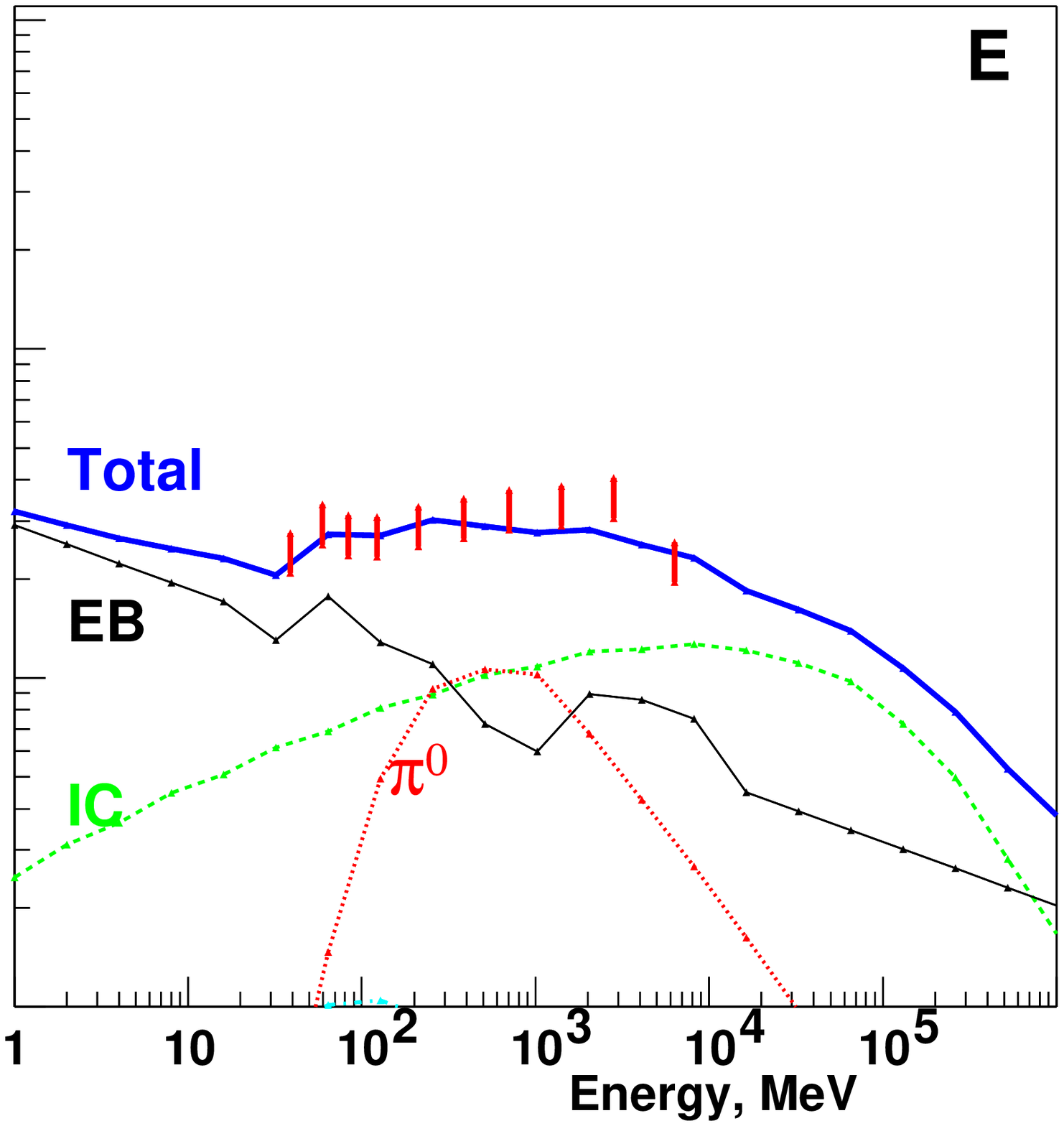}\hfill
 \includegraphics[width=0.307\textwidth,height=0.307\textwidth]{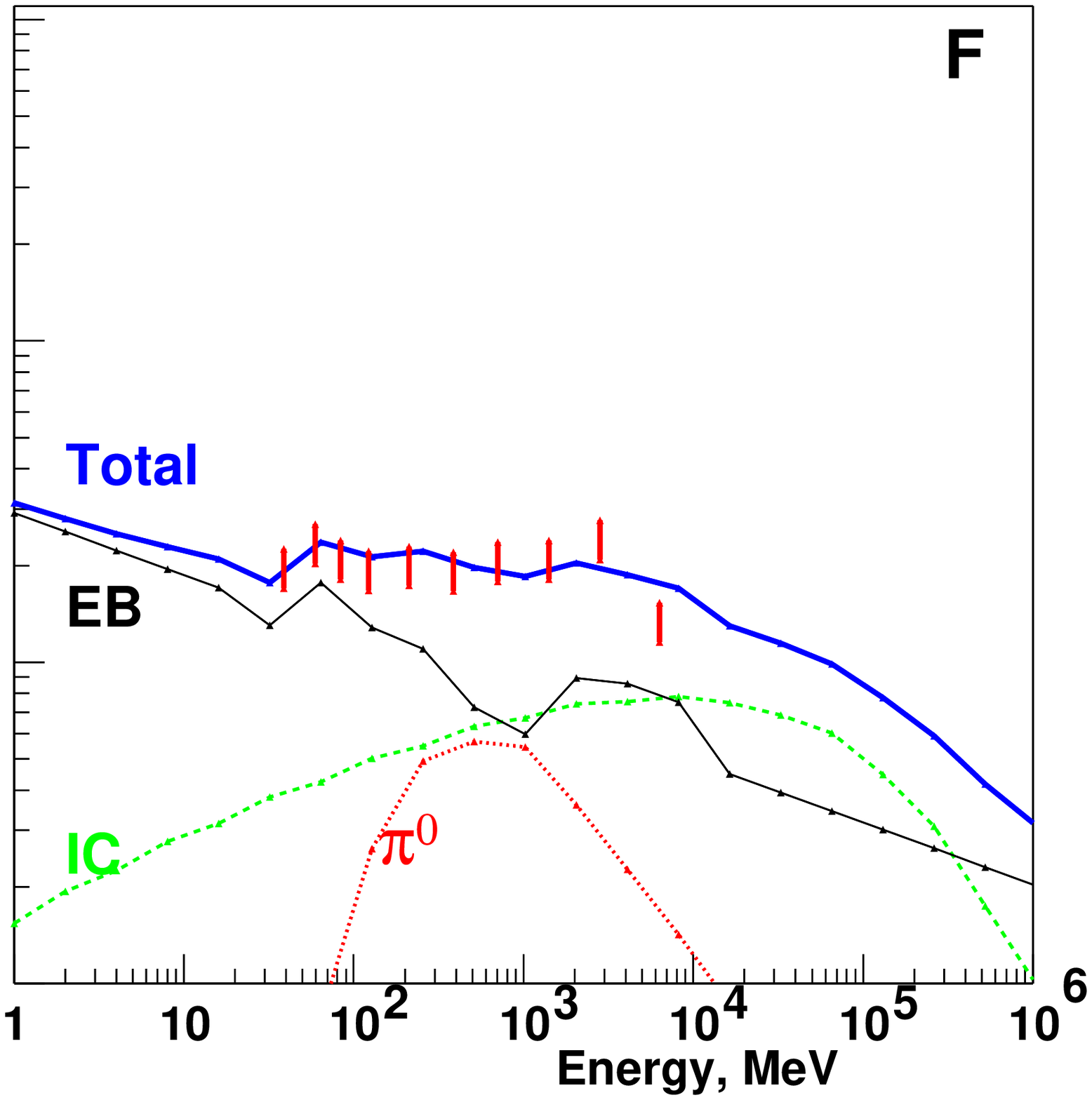}
  \end{center}
  \vspace{-0.5pc}
\caption{$\gamma$-ray spectra for optimized model. Regions as Fig.\ 1.}
\label{spectrum_500026}
\end{figure}

I.V.M.\ acknowledges partial support from
a NASA Astrophysics Theory Program grant.


\nobreak
\section{References}
\re
1.\ Berezhko E.G., V\"olk H.J.\ 2000, ApJ 540, 923
\re 
2.\ Moskalenko I.V., Strong A.W., Ormes J.F., Potgieter M.S.\ 2002, ApJ 565, 280
\re 
3.\ Strong A.W., Moskalenko I.V., Reimer O.\ 2000, ApJ 537, 763 (err.\ 541, 1109)
\re  
4.\ Strong A.W., Moskalenko I.V., Reimer O.\ 2003, these Proceedings
\endofpaper
\end{document}